\documentclass[12pt]{iopart}

\usepackage{graphicx}

\def\lesssim{{\
\lower-1.2pt\vbox{\hbox{\rlap{$<$}\lower5pt\vbox{\hbox{$\sim$}}}}\ }} 
\def\gtrsim{{\
\lower-1.2pt\vbox{\hbox{\rlap{$>$}\lower5pt\vbox{\hbox{$\sim$}}}}\ }}

\begin{document}

\title{Early reionization by decaying particles in light of three year WMAP data}

\author{Shinta Kasuya$^a$ and Masahiro Kawasaki$^b$}

\address{
$^a$ Department of Information Science,
     Kanagawa University, Kanagawa 259-1293, Japan\\
$^b$ Institute for Cosmic Ray Research,
     University of Tokyo, Chiba 277-8582, Japan}
\ead{kasuya@kanagawa-u.ac.jp, kawasaki@icrr.u-tokyo.ac.jp}

\begin{abstract}
We study the reionization histories where ionizing UV photons are emitted from decaying 
particles, in addition to usual contributions from stars and quasars, taking account of the fact 
that the universe is not fully ionized until $z \sim 6$ as observed by Sloan Digital Sky Survey.
Likelihood analysis of the three-year data from the WMAP (Wilkinson Microwave Anisotropy 
Probe) severely constrains the decaying particle scenario.In particular, the decaying particle 
with relatively short lifetime is not favored by the polarization data.
\end{abstract}


\maketitle


\section{Introduction}

Ionization history after recombination was not known for many years, until recent 
development of observations. The Sloan Digital Sky Survey (SDSS) obtained the first 
evidence of the reionization epoch \cite{SDSS}. The Gunn-Peterson test for the observed 
spectra of quasars revealed that there existed neutral hydrogen at the redshift $z\sim 6$, 
which implies that the completion of the reionization of the universe was seen at $z \sim 6$, 
although the amount of the neutral hydrogen at $z\gtrsim 6$ may not be very large to explain 
the observed Gunn-Peterson trough.

On the other hand, the Wilkinson Microwave Anisotropy Probe (WMAP) observation found the 
earlier reionization than $z\sim 6$ in terms of the large optical depth \cite{WMAP1,WMAP3}.
The first-year data favored very large optical depth $\tau_{op} = 0.17 \pm 0.04$ \cite{WMAP1}.
The conventional reionization by UV photons emitted from usual early formed stars and 
quasars, could only reach to the optical depth of $\sim 0.05$~\cite{LB,stars3} if it is consistent 
with the SDSS data. Of course, using the nonstandard initial mass function and 
unrealistic escaping UV fraction from galaxies, one could achieve a large optical depth 
\cite{stars1}. However, the UV fluxes from stars and quasars increase with time and full 
reionization takes place before $z\sim 6$. Thus, one has to consider a more complex ionization 
history which may be realized, for example, by the use of different photon emission processes 
between Population II and III stars~\cite{stars2}. 

An alternative line is to consider another source of UV photons. It could be decaying particles 
\cite{KKS,KK, decay}. In Refs.~\cite{KKS,KK}, we considered decaying particles in addition 
to conventional stars and quasars and found that the decaying particle scenario is consistent 
with the first year WMAP data for large parameter space.

Recently, WMAP team released three year data, in which the value of the optical depth is 
lowered: $\tau_{op} = 0.09 \pm 0.03$ \cite{WMAP3}. It changed towards the conventional
value of $\sim 0.05$, but some extra UV photons are still necessary. It would be possible
to have such a small amount of extra UV photons before $z\sim 6$ using less unrealistic 
assumptions for stars and quasars. In this article, however, we stick to the decaying particle
scenario to assess its capability. 

The structure of the paper is as follows. In the next section, we explain how we follow the 
evolution of the ionization history including the decaying particles. In Sect.~III, we evaluate
the $\chi^2$ for each model parameters, and construct likelihood contours for the parameters of
decaying particle scenario. At the same time, the detail of the power spectra of temperature
and polarization anisotropies in the cosmic microwave background (CMB) radiation are shown, 
and we discuss how it discriminates the preferred model parameters in the same section. The 
final section is devoted to our conclusions.

\section{Reionization history}
We consider that ionizing UV photons are emitted from decaying particles as well as from
usual stars (and quasars). The latter are indeed responsible for full ionization at $z\sim 6$. 
On the other hand, UV photons from particle decays keep a nonzero but small amount of
the ionization fraction before $z\sim 6$.

We follow the thermal history from $z>10^3$ including the recombination epoch, calculating the 
ionization fraction of hydrogen and helium, and the electron temperature, on the basis of the 
hierarchical clustering scheme of the cold dark matter scenario used in Fukugita and Kawasaki 
\cite{FK94}.

In addition, we include the contribution of UV photons from decaying particles \cite{KKS,KK}.
We assume that the particle $\phi$ emits two photons with monochromatic energy of half the 
mass of that particle, i.e., $E_\gamma =m_\phi/2$. The number density of $\phi$ particle is 
given by
\begin{equation}
n_\phi = n_\phi(0)(1+z)^3 e^{-t/\tau_\phi},
\end{equation}
where $\tau_\phi$ is the lifetime of the $\phi$ particle. Hydrogen atoms can be ionized if
the emitted photons have energy with $E_\gamma > 13.6$ eV. Then the source term for
the decaying particle can be written as
\begin{equation}
\left(\frac{dn_\gamma}{dt}\right)_{dp} = \frac{n_\phi}{\tau_\phi}.
\end{equation}
The amount of emitted photons is determined once the mass, $m_\phi$, lifetime, 
$\tau_\phi$, and abundance, $\Omega_\phi$, of the particle are fixed. We calculate the
ionization histories for $E_\gamma=15 - 1000$ eV and $\tau_\phi = 10^{14} - 10^{18}$ sec,
adjusting the abundance, $\Omega_\phi$, to get the optical depth in the range of
$\tau_{op} =0.075 - 0.2$, which is defined by
\begin{equation}
\tau_{op} = \int_0^\infty dz \ \sigma_T \left(\frac{dt}{dz}\right) \left[ n_e - n_e^{(sr)}\right],
\end{equation}
where $\sigma_T$ is the Thomson coss section and $n_e^{(sr)}$ is the electron number density 
for standard recombination. We subtract this term to estimate only the effect of reionization.

\section{Cosmic microwave background radiation}
As well known, the temperature and polarization anisotropies of the CMB radiation could inform 
us of what happened during the dark age after recombination. We now look for whether ionization 
histories in the decaying particle scenario are consistent with CMB observation in the three-year 
WMAP data.

We obtain the power spectra by the code modified from CMBFAST \cite{cmbfast} so as to 
equip us with the capability of using the evolution obtained for the ionization fraction and matter 
temperature, and evaluate $\chi^2$ using the code provided by WMAP \cite{WMAP3}.
We seek for the range $E_\gamma=15 -1000$ eV, $\tau_\phi=10^{14} - 10^{18}$ sec (with 9 
bins), and $\tau_{op}=0.075 - 0.2$ (with 6 bins). For fixed mass, lifetime, and optical depth,  
$\chi^2$ is calculated adjusting $\Omega_b$, $\Omega_m$, $H_0$, $n_s$, and the amplitude 
of the spectrum. We construct the contour of likelihood with grid-based analysis, using 
$\Delta\chi^2 = \chi^2 -\chi^2|_{\rm best}$, where $\chi^2|_{\rm best}$ is the six-parameter 
$\chi^2$ minimum for $\Lambda$CDM with WMAP-only.

Before going into the analyses with hybrid UV source of stars and decaying particles, let us ask 
a question: Is it possible that the decaying particles alone could provide an ionization history 
consistent with SDSS and three-year WMAP? It is indeed possible to have an ionization 
history such that the ionization fraction rises rapidly at $z \sim 6$, which is consistent with 
SDSS observation. For example, this is realized with $\tau_\phi = 9.5 \times 10^{17}$ sec, 
$\Omega_\phi=1.2\times 10^{-5}$, and $m_\phi=80$ eV. However, the $\chi^2$ is about 
$4\sigma$ larger than that of the best-fit $\Lambda$CDM model. Of course, the optical depth 
has a very large value, such as 0.23. Thus, we can exclude this possibility.

We next consider another limiting case, that the ionization fraction becomes unity at 
$z\sim 6$ due to UV photons emitted from stars and quasars only, and there is no contribution 
from decaying particles, which leads to the optical depth $\tau_{op}\approx 0.05$. In this case,
the $\chi^2$ is just outside the  $1\sigma$ region. One may hope that the $\chi^2$
will be reduced once an extra contribution of UV photons from decaying particles is included. As 
we shall see below, however, this is not the case.

\begin{figure}[h]
\includegraphics[width=80mm]{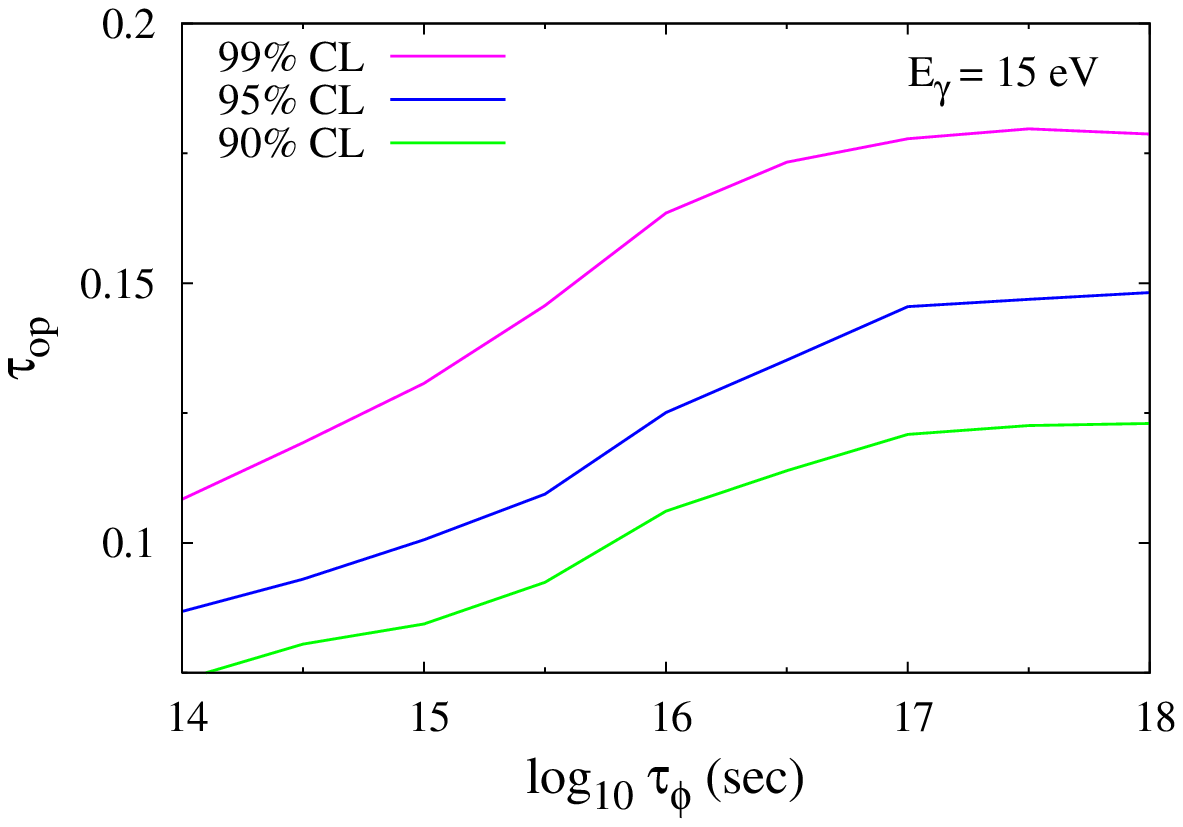}
\includegraphics[width=80mm]{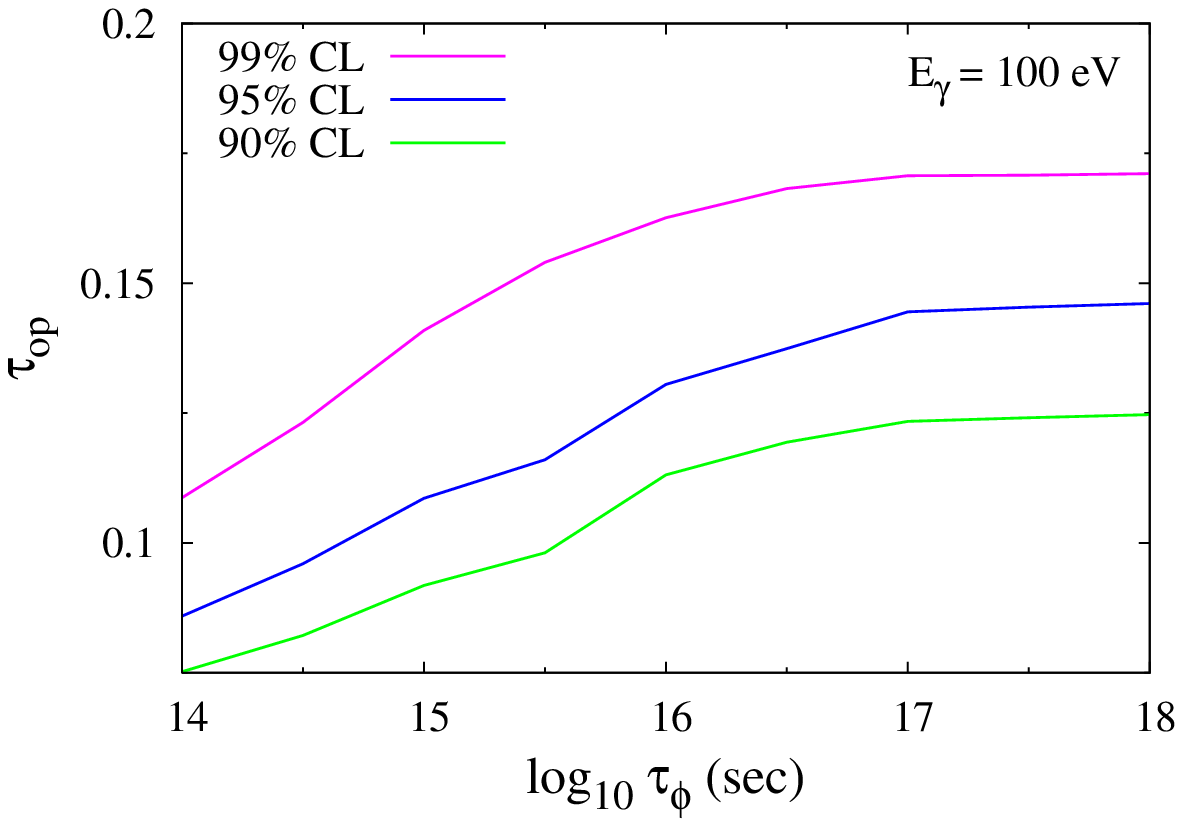}
\begin{center}
\includegraphics[width=77mm]{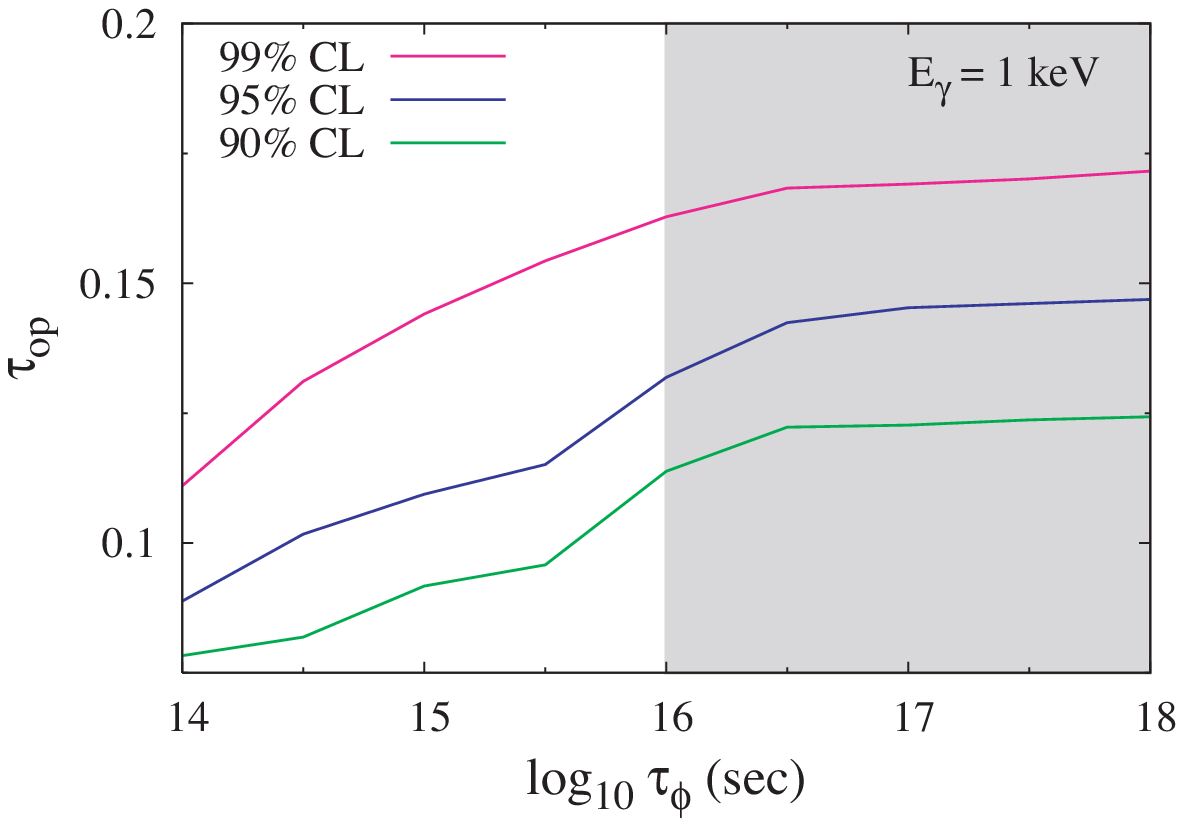}
\end{center}
\caption{Likelihood contours of the decaying particle scenario with $E_\gamma=15$ eV (top-left),
100 eV (top-right), and 1 keV (bottom). We show 90, 95, and 99\% CL lines. Lifetime longer 
than $\sim 10^{16}$ sec (shaded region)  is excluded by diffuse X-ray
background observation in the bottom panel.
\label{contours}}
\end{figure}

We show the likelihood contours for $E_\gamma = 15$, 100, and 1000 eV, in the top-left, 
top-right, and bottom panels, respectively, in Fig.~\ref{contours}. They look very similar to 
each other. Contrary to our expectation, there is no $1\sigma$ region. We can see that there 
is a weak dependence on the lifetime, and that the larger optical depth is not favored.

\begin{figure}[h]
\begin{center}
\includegraphics[width=78mm]{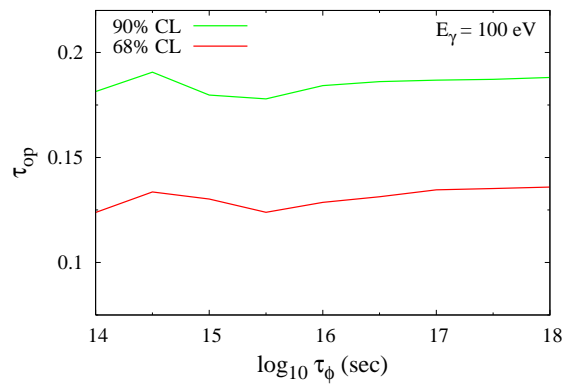}
\end{center}
\caption{Likelihood contours of the decaying particle scenario with $E_\gamma=100$ eV,
using only temperature anisotropy data. We show 68 and 90\% CL lines.
\label{cont_TT}}
\end{figure}

It is wise to compare these with the contours for which only the temperature power spectrum is 
used for estimating the likelihood. We can then see that the $\chi^2$ does not depend on the 
lifetime $\tau_\phi$ at all, as can be seen in Fig.~\ref{cont_TT} for $E_\gamma=100$ eV.
Moreover, there is a $1\sigma$ region: $\tau_{op} \lesssim 0.13 (0.17)$ is allowed for the 
68\% (90\%) confidence level.  This is easy to understand, because the temperature power 
spectrum is dependent on only the value of the optical depth, and does not `see' the actual 
ionization history. On the other hand, the polarization anisotropy in the CMB radiation 
discriminates among the different ionization histories. This is why we can see the dependence 
on the lifetime in Fig.~\ref{contours}, and the longer lifetime seems to be preferable.

Notice that lifetime longer than $10^{16}$ sec for $E_\gamma= 1$ keV is excluded. This is
because the flux of the photons exceeds the observed diffuse X-ray background \cite{KKS},
although the ionization history respects the Gunn-Peterson trough observed by SDSS.

\begin{figure}[h]
\begin{center}
\includegraphics[width=80mm]{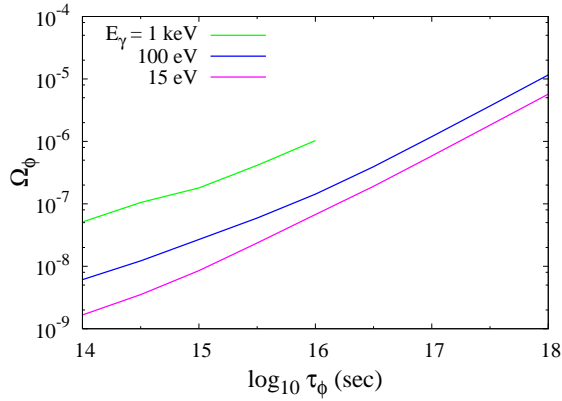}
\end{center}
\caption{Allowed region of the parameter space of the decaying particle scenario. The lines are
drawn for the 99\% CL contours in Fig.~\ref{contours}, which are thus the upper limits for the 
amount of the decaying particles, $\Omega_\phi$, for fixed lifetime $\tau_\phi$.
\label{para_sp}}
\end{figure}

We can thus obtain the allowed region for the decaying particle scenario in the parameter
space $(\tau_\phi, \Omega_\phi)$. We show each line with a 99\% confidence level in 
Fig.~\ref{para_sp}. For $E_\gamma \lesssim 100$ eV, longer lifetime is also allowed, until the 
abundance reaches the whole amount of the dark matter, since it is the ratio of 
$n_\phi/\tau_\phi$ that determines the amount of the emitted photons from the decaying 
particles for lifetime longer than the age of the universe. Among the decaying
particles, saxion, the scalar partner of the axion, in the gauge-mediated supersymmetry 
breaking scenario, may be the best candidate. Its lifetime is about $10^{17}$ sec for
the saxion mass $m_s \sim 100$ eV and the axion decay constant $F_{PQ}\sim 10^{12}$ GeV.
Since the saxion decays into photons through a one-loop diagram with branching ratio of order 
$\sim 10^{-7} - 10^{-6}$, it matches to the allowed region of the parameter space very well. 
Notice that these constraints apply to most kinds of decaying particles which emit photons
with branching ratio $B_\gamma$ provided that one regards the effective abundance as 
$B_\gamma \Omega_\phi$. 

\begin{figure}[h]
\includegraphics[width=80mm]{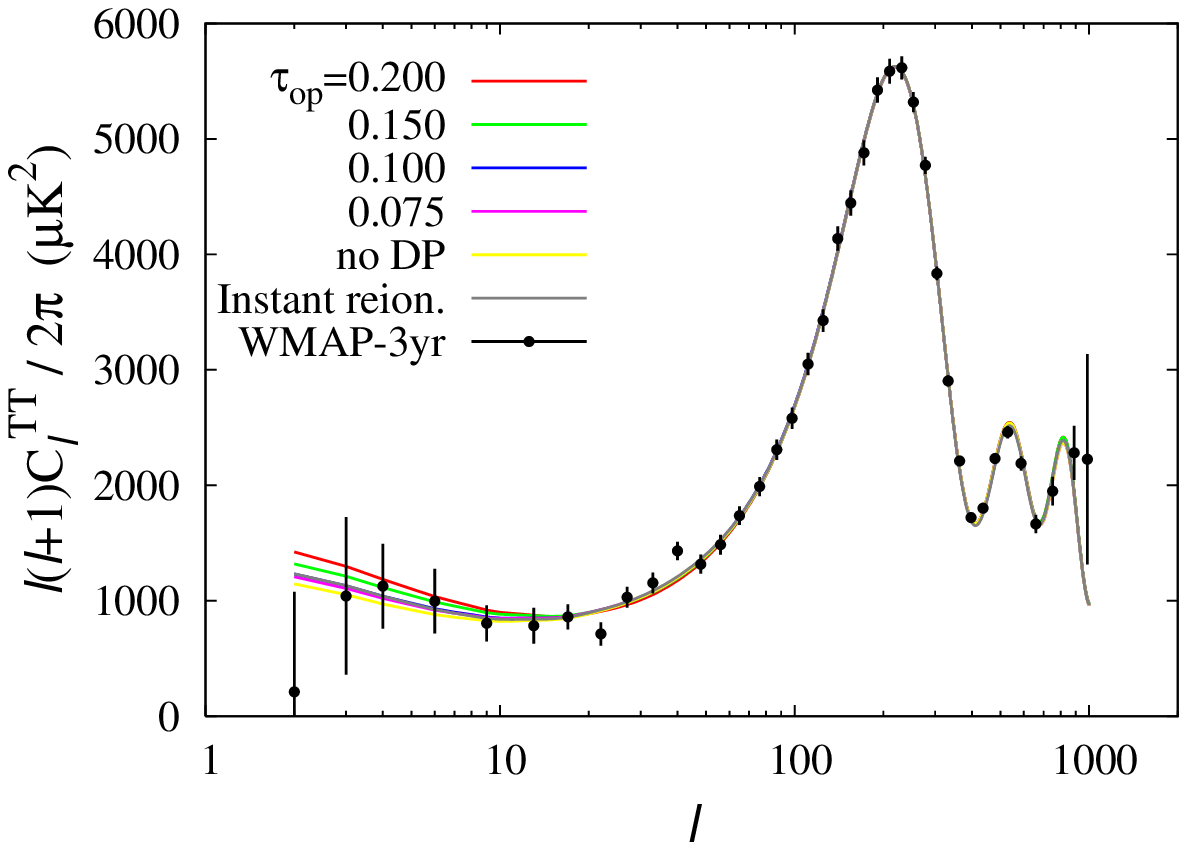}
\includegraphics[width=80mm]{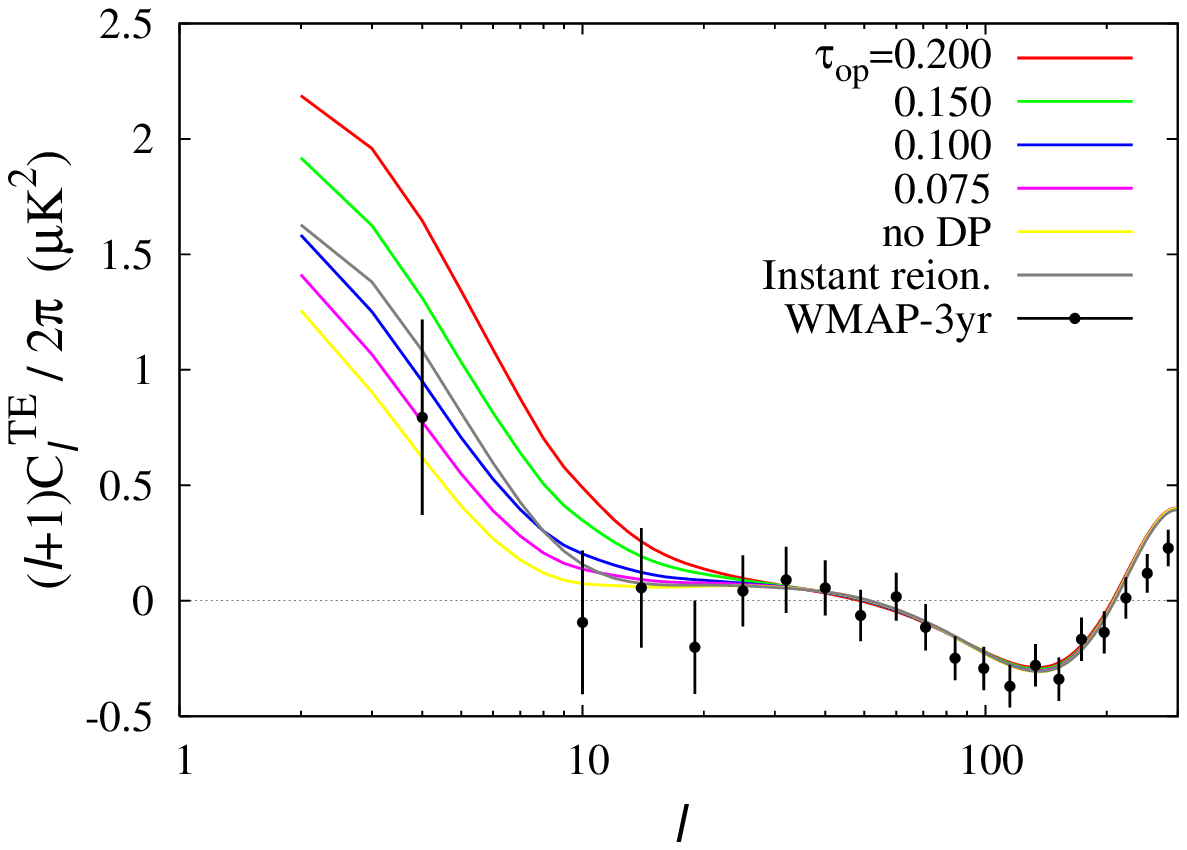}
\begin{center}
\includegraphics[width=80mm]{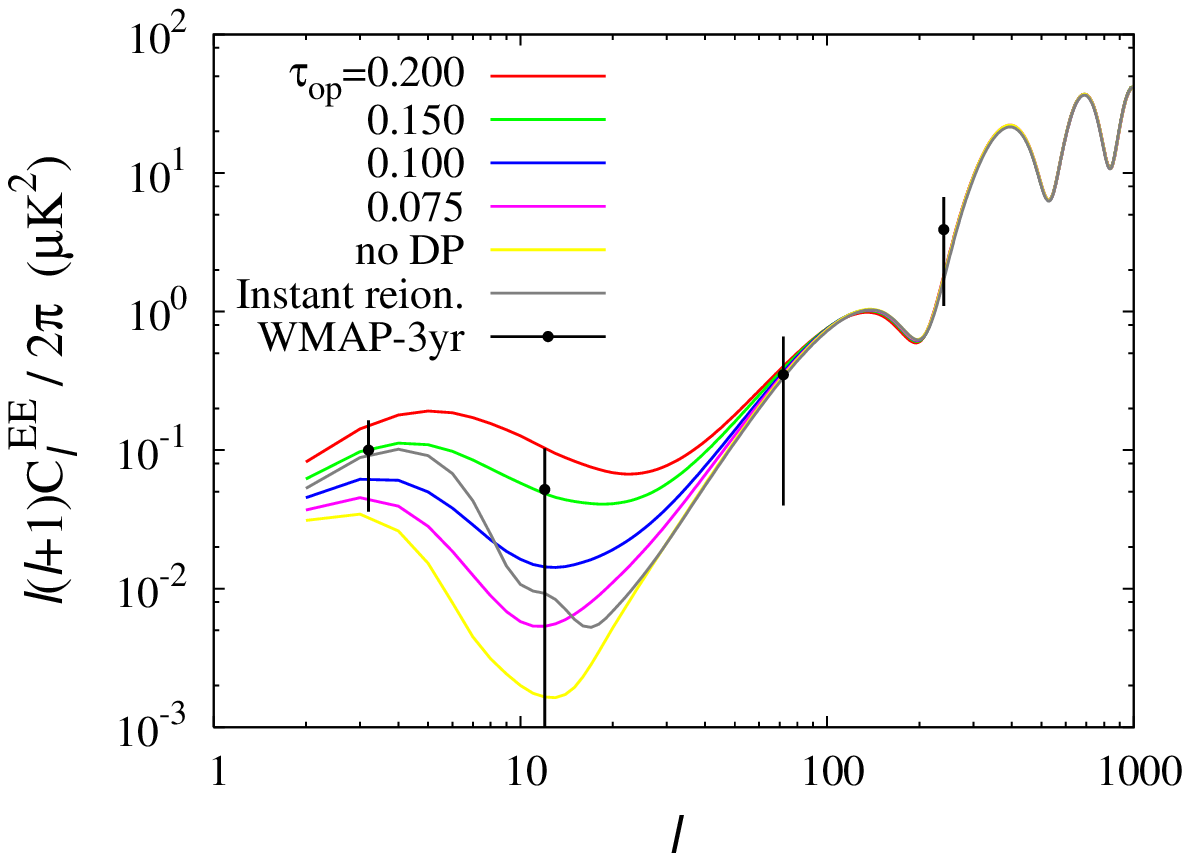}
\end{center}
\caption{TT (top-left), TE (top-right), and EE (bottom) power spectra of the CMB for 
$\tau_\phi=10^{17}$ sec with various optical depths. We also show the spectra
for the limiting case of no decaying particle contribution, and the (one-step) instantaneous 
reionization. The three year WMAP data are plotted as well. 
\label{life17}}
\end{figure}

Now let us move on to the detailed power spectra of temperature and polarization anisotropies
in CMB. Temperature (TT), temperature-polarization (TE), and polarization (EE) power spectra
are shown for optical depth in the range of $\tau_{op}=0.075 - 0.2$ with fixed lifetime 
$\tau_\phi=10^{17}$ sec in Fig.~\ref{life17}. For the limiting case of the decaying particle 
scenario, we also plot the spectra without any contributions from decaying particles 
($\tau_{op} \approx 0.05$). In addition, one-step instantaneous reionization case 
($\tau_{op} \approx 0.09$) is shown.

Since it is not easy to obtain some information only from the figures of the spectra, we give
the breakdown of the $\chi^2$ in these cases in upper half of Table~\ref{t-chi2}. Here, we 
tabulate the difference of $\chi^2$ from the minimum case, $\Delta\chi^2$, defined above.
Larger optical depth is disfavored by all the components, which may be seen in all the spectra
in Fig.~\ref{life17}. The case with $\tau_{op} \sim 0.01$ seems to be slightly favored by the 
polarization data, which almost mimic the instantaneous reionization case in TE spectra. 
However, the significance is not so great, and TT data do not prefer this case. Anyway, the 
total $\chi^2$ slightly favours smaller optical depth.

\begin{table}[h]
\caption{$\Delta\chi^2$ for various optical depth for $\tau_\phi=10^{17}$ sec (upper), and
for various lifetimes for $\tau_{op}\approx 0.1$ (lower). In the upper half, the last line 
represents for the case without contributions from decaying particles. TT, low TT, TE, and low 
pol. stand for the TT power spectrum, pixel based TT for low $\ell$, TE power spectrum, and
pixel based polarization for low $\ell$, respectively.
\label{t-chi2}}
\vspace{2mm}
\begin{center}
\begin{tabular}{ccrrccr}
\hline\hline
$\tau_\phi$ \ \ & $\tau_{op}$ \ & \multicolumn{1}{c}{\ TT} & \multicolumn{1}{c}{low TT} 
& TE & \ low pol. \ & \multicolumn{1}{c}{tot.} \\
\hline
$10^{17}$ \ &  0.199 \ &  3.34 \ &  1.48  \ \  &  1.68 &  6.62 &   13.13 \\
$10^{17}$ \ &  0.150 \ &  1.81 \ &  0.78 \ \  &  0.94 &  2.83 &    6.37 \\
$10^{17}$ \ &  0.100 \ &  1.06 \ &  0.10 \ \   &  0.33 &  1.60 &    3.10 \\
$10^{17}$ \ &  0.074 \ & $-0.07$ \ &  0.03 \ \   &  0.76 &  1.93 &    2.66 \\
    ---        \ &  0.050 \ & $-0.24$ \ & $-0.23$ \ \ & 0.44    &  2.44  &   2.42 \\
\hline
$10^{18}$ \ &  0.100 &  1.05 \ &  0.10 \ \   &  0.32 &  1.56 &    3.04 \\
$10^{17}$ \ &  0.100 &  1.06 \ &  0.10 \ \  &  0.33 &  1.60 &    3.10 \\
$10^{16}$ \ &  0.100 &  1.24 \ &  0.10 \ \  &  0.36 &  1.93 &    3.58 \\
$10^{15}$ \ &  0.101 &  1.16 \ &  0.01 \ \  &  1.42 &  2.65 &    5.25 \\
$10^{14}$ \ &  0.100 &  1.21 \ & $-0.23$ \ \  &  4.04 &  2.79 &    7.82 \\
\hline\hline
\end{tabular}
\end{center}
\end{table}

\begin{figure}[h]
\includegraphics[width=80mm]{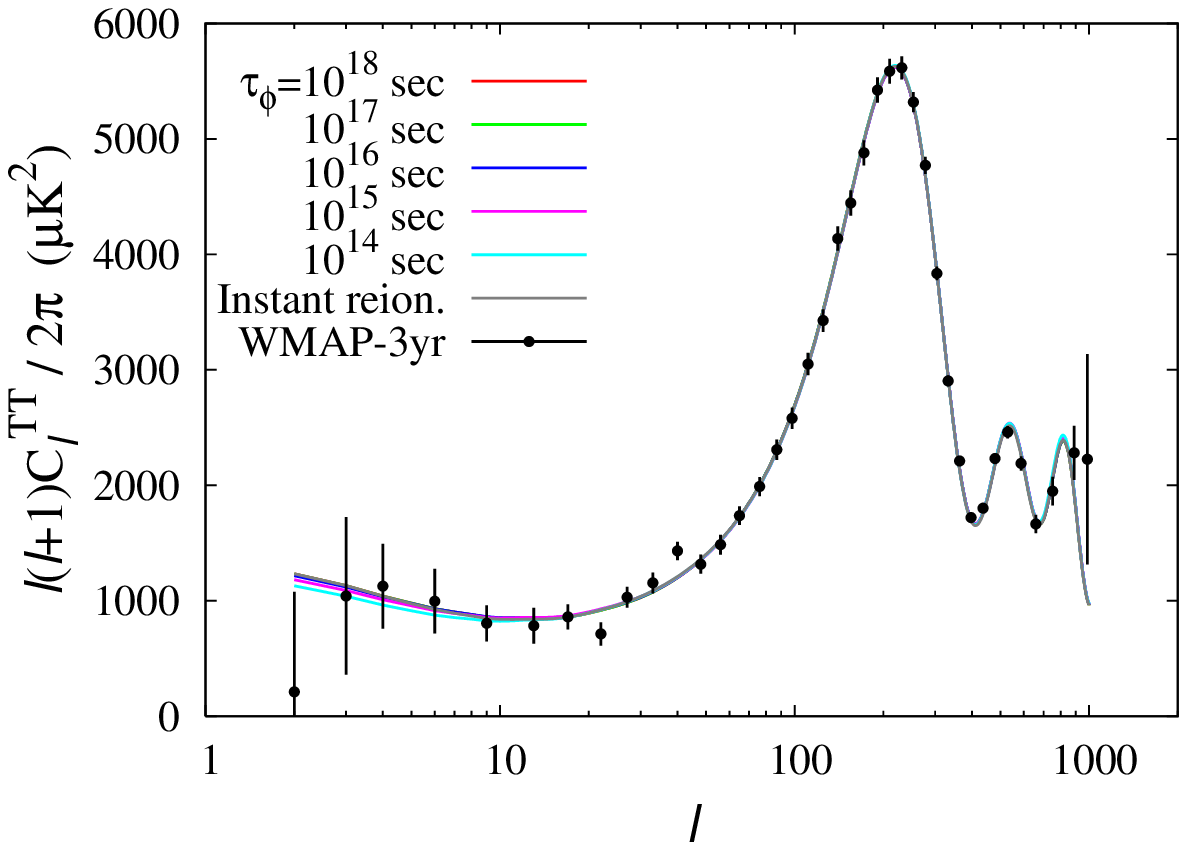}
\includegraphics[width=80mm]{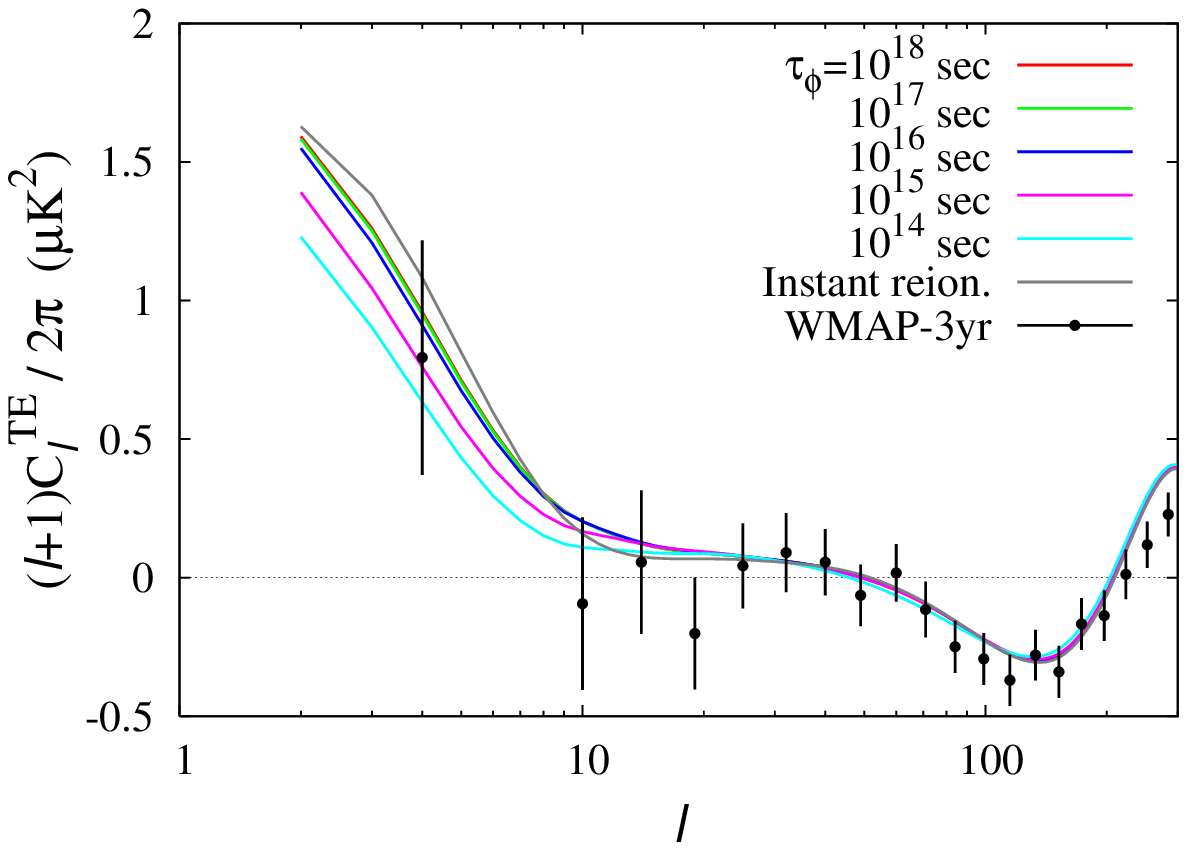}
\begin{center}
\includegraphics[width=80mm]{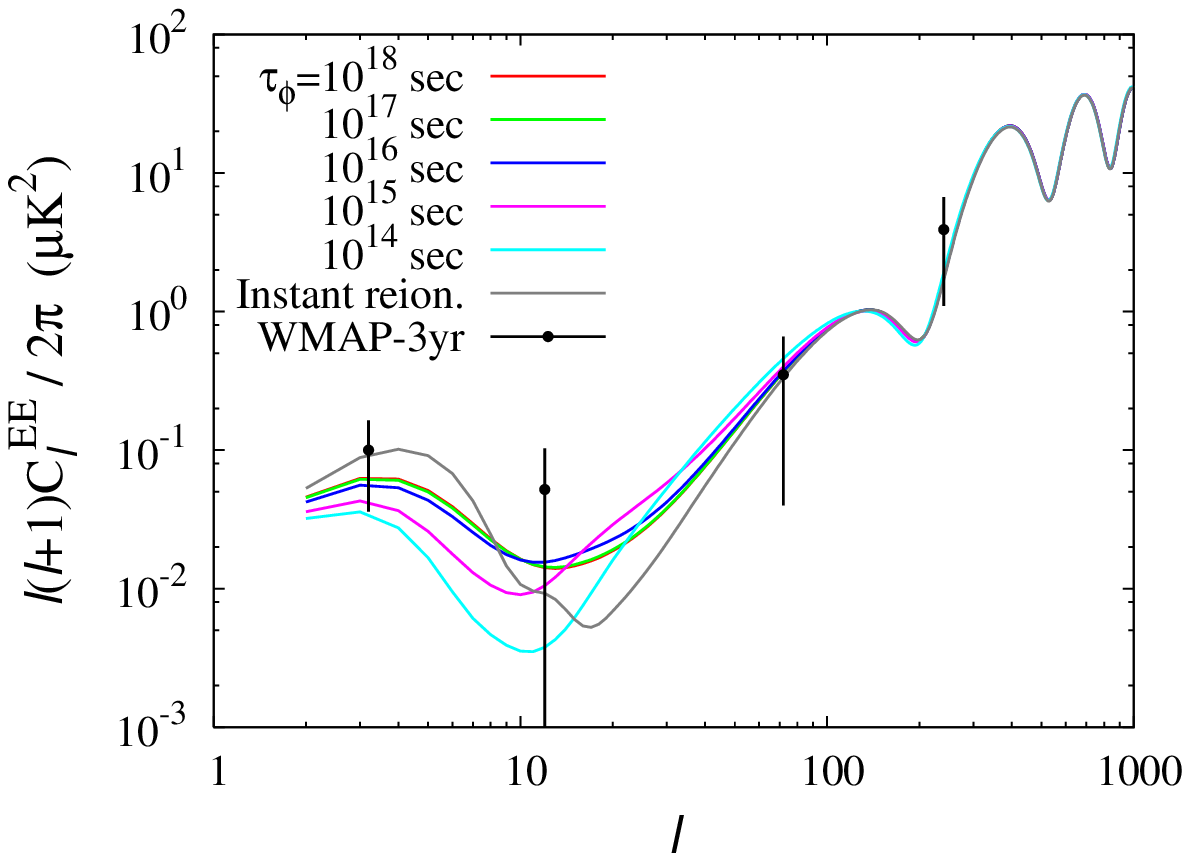}
\end{center}
\caption{TT (top-left), TE (top-right), and EE (bottom) power spectra of the CMB for 
$\tau_{op}=0.1$with various lifetimes. Also shown are the spectra for the (one-step) 
instantaneous reionization, and the three-year WMAP data.
\label{opt01}}
\end{figure}

On the other hand, one can look for the dependence on the lifetime with fixed optical depth
in Fig.~\ref{opt01} and the lower half of Table~\ref{t-chi2}. In this case, TT data do not
favour any lifetime, since the TT power spectra is almost only sensitive to the value of the 
optical depth, not to the history of actual ionization. Short lifetime is severely disfavored since
it has large $\chi^2$ from the polarization data. This is not so apparent from the figures for
either TE or EE power spectra, but it seems that larger power is needed in the low $\ell$
regions in order to fit the polarization data much better.

\section{conclusions}
We have considered the ionizing UV photons emitted from decaying particles, in addition to
the usual contributions of stars and quasars. The stars and quasars are responsible for the 
quick rise of the inonization fraction at $z\sim 6$ observed by SDSS as the Gunn-Peterson 
trough. On the other hand, the decaying particles cause partial ionization at $z > 6$.

The decaying particle scenario could explain the large optical depth, and was consistent 
with the first-year WMAP data. However, the three-year WMAP data do not favour the 
decaying particle scenario so much. Notice that our results apply to most kinds of decaying 
particles which emit photons with branching ratio $B_\gamma$ provided that one regards 
the effective abundance as $B_\gamma \Omega_\phi$.

The ionization by the decaying partcles with $\tau_{op} \lesssim 0.17$ does indeed seem to be 
consistent within $2\sigma$ with TT power spectrum. The peculiar aspect of the three-year
WMAP data is in the polarization; in particular, the polarization data do not allow the 
earlier reionization which is a distinct feature of the ionization history from the 
UV emission from particle decays, and hence the longer lifetime is favoured. 
Since more power at low $\ell$ is favoured by the polarization anisotropy, one may 
ambitiously claim that a better fit may be obtained for the decaying particle scenario with 
increasing decay rate.

\section*{Acknowledgments}
S.K. is grateful to Kazuhide Ichikawa for useful discussions.
The work of S.K. is supported by the Grant-in-Aid for Scientific Research from the
Ministry of Education, Science, Sports, and Culture of Japan, No.~17740156.




\section*{References}
\begin{thebibliography}{10}

\bibitem{SDSS}   X. Fan {\it et al.},
  Astron.\ J.\  {\bf 123}, 1247 (2002).
  

\bibitem{WMAP1} 
A.~Kogut {\it et al.},
  Astrophys.\ J.\ Suppl.\  {\bf 148}, 161 (2003);
  D.~N.~Spergel {\it et al.} ,
  Astrophys.\ J.\ Suppl.\  {\bf 148}, 175 (2003).
  

\bibitem{WMAP3}
D.~N.~Spergel {\it et al.},
astro-ph/0603449;
L.~Page {\it et al.},
astro-ph/0603450;
G.~Hinshaw {\it et al.},
  astro-ph/0603451;
N.~Jarosik {\it et al.},
  astro-ph/0603452.

\bibitem{LB}
  A.~Loeb and R.~Barkana,
  Ann.\ Rev.\ Astron.\ Astrophys.\  {\bf 39}, 19 (2001).

\bibitem{stars3}
W.~A.~Chiu, X.~Fan and J.~P.~Ostriker,
  Astrophys.\ J.\  {\bf 599}, 759 (2003);
C.~A.~Onken and J.~Miralda-Escud\'e,
  Astrophys.\ J.\  {\bf 610}, 1 (2004).



\bibitem{stars1}      
M.~Fukugita and M.~Kawasaki,
  Mon.\ Not.\ Roy.\ Astron.\ Soc.\  {\bf 343}, L25 (2003);
B.~Ciardi, A.~Ferrara and S.~D.~M.~White,
  Mon.\ Not.\ Roy.\ Astron.\ Soc.\  {\bf 344}, L7 (2003);
 X.~L.~Chen, A.~Cooray, N.~Yoshida and N.~Sugiyama,
  Mon.\ Not.\ Roy.\ Astron.\ Soc.\  {\bf 346}, L31 (2003).

  
\bibitem{stars2} 
R.~Cen,
  Astrophys.\ J.\  {\bf 591}, 12 (2003);
L.~Hui and Z.~Haiman,
  Astrophys.\ J.\  {\bf 596}, 9 (2003);
A.~Sokasian, T.~Abel, L.~Hernquist and V.~Springel,
  Mon.\ Not.\ Roy.\ Astron.\ Soc.\  {\bf 344}, 607 (2003);
A.~Sokasian, N.~Yoshida, T.~Abel, L.~Hernquist and V.~Springel,
  Mon.\ Not.\ Roy.\ Astron.\ Soc.\  {\bf 350}, 47 (2004);
A.~J.~Benson, N.~Sugiyama, A.~Nusser and C.~G.~Lacey,
  astro-ph/0512364.


 
\bibitem{KKS} S.~Kasuya, M.~Kawasaki and N.~Sugiyama,
Phys.\ Rev.\ D {\bf 69}, 023512 (2004).
  
\bibitem{KK}
S.~Kasuya and M.~Kawasaki,
Phys.\ Rev.\ D {\bf 70}, 103519 (2004).


\bibitem{decay} 
S.~H.~Hansen and Z.~Haiman,
  Astrophys.\ J.\  {\bf 600}, 26 (2004);
X.~L.~Chen and M.~Kamionkowski,
  Phys.\ Rev.\ D {\bf 70}, 043502 (2004);
P.~P.~Avelino and D.~Barbosa,
  Phys.\ Rev.\ D {\bf 70}, 067302 (2004);
P.~L.~Biermann and A.~Kusenko,
  Phys.\ Rev.\ Lett.\  {\bf 96}, 091301 (2006);
M.~Mapelli, A.~Ferrara and E.~Pierpaoli,
  Mon.\ Not.\ Roy.\ Astron.\ Soc.\  {\bf 369}, 1719 (2006);
L.~Zhang, X.~L.~Chen, Y.~A.~Lei and Z.~G.~Si,
  astro-ph/0603425;
Y.~A.~Shchekinov and E.~O.~Vasiliev,
  astro-ph/0604231.


\bibitem{FK94} 
M.~Fukugita and M.~Kawasaki,
  Mon. Not. R. Astron. Soc. {\bf 269}, 563 (1994).

\bibitem{cmbfast} 
U.~Seljak and M.~Zaldarriaga,
  Astrophys.\ J.\  {\bf 469}, 437 (1996).



\end{thebibliography}
\end{document}